\documentclass{cimento}
\usepackage{lipsum}
\usepackage{subfigure}

\title{Measuring $\mu_B$ at the LHC with ALICE via antiparticle-over-particle ratios}
\author{M.~Ciacco\from{ins:x}\thanks{mario.ciacco@cern.ch}, on behalf of the ALICE Collaboration}
\instlist{\inst{ins:x} Dipartimento DISAT del Politecnico and Sezione INFN - Turin, Italy}
\usepackage{amsmath}
\usepackage{graphicx}  % got figures? uncomment this

\usepackage{lineno}
\begin{document}

\maketitle

%\linenumbers

\begin{abstract}
The baryon chemical potential $\mu_B$ is a fundamental parameter for the statistical mechanical description of particle production in heavy-ion collisions: its value encodes the average net baryon content of the fireball at chemical freeze-out. The first $\mu_B$ measurement in high-energy Pb-Pb collisions at the LHC was published in 2018 in Nature, and it was found that $\mu_B$=0.7$\pm$3.8 MeV. In this contribution, an improved $\mu_B$ measurement based on the study of antiparticle-to-particle ratios for protons, $^3\mathrm{He}$ and hypertriton ($^3_\Lambda\mathrm{H}$) using the data collected by ALICE in Run 2 of the LHC, is presented. The isospin chemical potential is also determined via the $\pi^-/\pi^+$ ratio. The obtained $\mu_B$ represents the most precise measurement available.
%  The study of the pion, proton and helium ratios is based on standard analyses, while the selection of $^3_\Lambda\mathrm{H}$ candidates, which are reconstructed via the two-body charged-$\pi$ mesonic decay of the hypertriton, is performed using Boosted Decision Trees
\end{abstract}

\section{Introduction}
In hadronic systems, the baryon chemical potential $\mu_B$ is defined as the variation of the system energy, $U$, caused by a change in its baryon content, $N_{B}$, i.e. $\mu_B=\partial U/\partial N_B$. This quantity is introduced in the Grand Canonical formulation of the Statistical Hadronisation Model (SHM) \cite{ref:SHM, ref:SHMVovchDonig, ref:SHMVovch}, a statistical mechanical description of the hadronic phase based on the assumption that the fireball can be described as a gas of hadrons and resonances in local equilibrium. The SHM has been extensively used to extract the temperature, $T_{\mathrm{ch}}$, and chemical potential at the so called ``chemical freeze-out'', i.e. the moment in which the inelastic interactions between the particles cease, from experimental data at different collision energies \cite{ref:PBMAGS, ref:PBMRHIC, ref:Rafelski}. From these comprehensive studies, it was observed that the baryon chemical potential smoothly decreases from $\mu_B\approx 1 \ \mathrm{GeV}$ at $\sqrt{s_{\mathrm{NN}}}\approx 1 \ \mathrm{GeV}$ to $\mu_B\approx 1 \ \mathrm{MeV}$ at the TeV scale: this trend is understood in terms of an increasing nuclear transparency for increasing collision energy, i.e. a progressively smaller fraction of the initial baryon number is transported to the interaction region. The first measurement of $\mu_B$ at the LHC was obtained by comparing the model predictions with the yields both of particles and antiparticles of various hadron species determined by the ALICE Collaboration in Pb-Pb collisions at a centre-of-mass energy per colliding nucleon of $\sqrt{s_{\mathrm{NN}}}=2.76\ \mathrm{TeV}$, in the 10\% most central events \cite{ref:DecodingQCD, ref:AndronicPLB}. It was found that the SHM accurately fits the data across nine orders of magnitude when $\mu_B=0.7\pm3.8\ \mathrm{MeV}$ and $T_{\mathrm{ch}}=156.2\pm1.5\ \mathrm{MeV}$. To obtain a more precise $\mu_B$ measurement, it is possible to consider antiparticle-to-particle ratios rather than the yields themselves \cite{ref:PhysRevC74}. From the experimental side, computing ratios allows us to cancel out systematic effects which are correlated between particles and their charge conjugates, thus reducing the uncertainty on the extracted SHM parameters. In the SHM, under the assumption of strangeness neutrality \cite{ref:ZPhys}, the ratios are connected to the baryon chemical potential as:
\begin{linenomath}
\begin{equation}\label{eq-1}
    R\propto\exp\left[-2\left(B+\frac{S}{3}\right)\frac{\mu_B}{T_{\mathrm{ch}}}-2\frac{\mu_{I_3}}{T_{\mathrm{ch}}}\right],
\end{equation}
\end{linenomath}
where $B$, $S$, and $I_{3}$ are the baryon number, strangeness, and isospin third component of the considered hadron species, respectively, while $\mu_{I_{3}}$ represents the isospin chemical potential. It can be observed that the ratios decrease exponentially with increasing baryon number. For this reason, we consider hadron species with a large baryon content, such as $^3\mathrm{He}$ and hypertriton $^3_{\Lambda}\mathrm{H}$ (a bound state of a proton, neutron, and $\Lambda$), both of which are characterised by $B=3$. In addition, the antiproton-over-proton ratio is considered in this work to increase the precision on $\mu_B$: indeed, in heavy-ion collisions, protons are the most copiously produced baryons. As a result, the SHM fit to ratios is driven primarily by the $\overline{\mathrm{p}}/\mathrm{p}$ ratio in the baryon sector. Charged pions are also considered in this work to precisely constrain the value of $\mu_{I_{3}}$, as they carry neither baryon number nor strangeness. From Eq.~\ref{eq-1}, it can be also obtained that, at the LHC energy, where $T_{\mathrm{ch}}\approx 155\ \mathrm{MeV}$ and $\mu_B\approx 1\ \mathrm{MeV}$, the temperature dependence of the antiparticle-to-particle ratios is negligible. In this work, $T_{\mathrm{ch}}$ is set to different values around a nominal temperature to study the systematic uncertainty induced by this parameter.

\section{The ALICE apparatus}
The ALICE apparatus is described in detail in Ref.~\cite{ref:ALICE08}. The main subdetectors used in this work are the V0, the Inner Tracking System (ITS), the Time Projection Chamber (TPC), and the Time-of-Flight (TOF). The V0 system \cite{ref:ALICE13} is composed of two arrays of plastic scintillators placed along the beam axis on both sides of the nominal interaction point, covering the $2.8<\eta<5.1$ and $-3.7<\eta<-1.7$ pseudorapidity intervals, respectively. The amplitude of the V0 signal is correlated with the charged-particle multiplicity, thus providing information about the centrality of the event, i.e. the displacement between the centres of the colliding nuclei in the plane transverse to the beam direction. The V0 signal is also used as a centrality trigger. The tracking of charged particles is performed using the ITS and TPC detectors. The ITS \cite{ref:ALICE10} is a six-layer silicon tracker located around the beam pipe. Due to its optimal space resolution of $O(10\ \mu\mathrm{m})$ in its two innermost layers, it provides a precise reconstruction of the track points close to the interaction point. In addition, the ITS information is used to determine the position of the primary interaction vertex with a resolution of about $100\ \mu\mathrm{m}$ along the beam axis. The ITS is surrounded by the TPC \cite{ref:ALICENIM}, a large gaseous detector. It constitutes the main tracking device of ALICE, as it reconstructs up to 159 space points for each particle crossing its full active volume. Thanks to its analog readout, it also measures the energy deposited by the particles in the gas, thus allowing for the identification of particle species by combining the specific energy loss $\mathrm{d}E/\mathrm{d}x$, obtained with a $5\%$ resolution, with the track momentum $p$. The particle identification (PID) capabilities of ALICE are extended to the intermediate momentum region by the TOF \cite{ref:ALICETOF}, which provides a precise measurement of the particle velocity $\beta$ by determining the particle flight time with a resolution better than $60\ \mathrm{ps}$. Moreover, the TOF information is used to determine the time of the collision. The ITS, TPC, and TOF cover the full azimuth and the pseudorapidity interval $\vert\eta\vert<0.8$, and they are placed in a uniform magnetic field of $0.5\ \mathrm{T}$ provided by a room-temperature solenoid.

\section{Data analysis}
The results reported in this contribution are obtained from the data sample collected by the ALICE collaboration during the 2018 LHC Pb-Pb run at $\sqrt{s_{\mathrm{NN}}}=5.02\ \mathrm{TeV}$. The trigger conditions are based on both the amplitude and timing of the V0 signal: in particular central and semicentral events are selected on top of the minimum bias baseline defined by a coincidence of signals in both V0 detectors. Additional offline selections are applied to reject events containing more than one reconstructed interaction vertex. The resulting data sample consists of about $3\times10^{8}$ events. Thanks to the large sample size it is possible to perform a centrality differential measurement of $\mu_B$: the centrality intervals considered in this work are 0-5\%, 5-10\%, and 30-50\%.
%%%% ADD INFO ABOUT TRACK SELECTIONS

\begin{figure}[!]
\centering
\begin{subfigure}{}
    \centering
    \includegraphics[height=0.6\textwidth,clip]{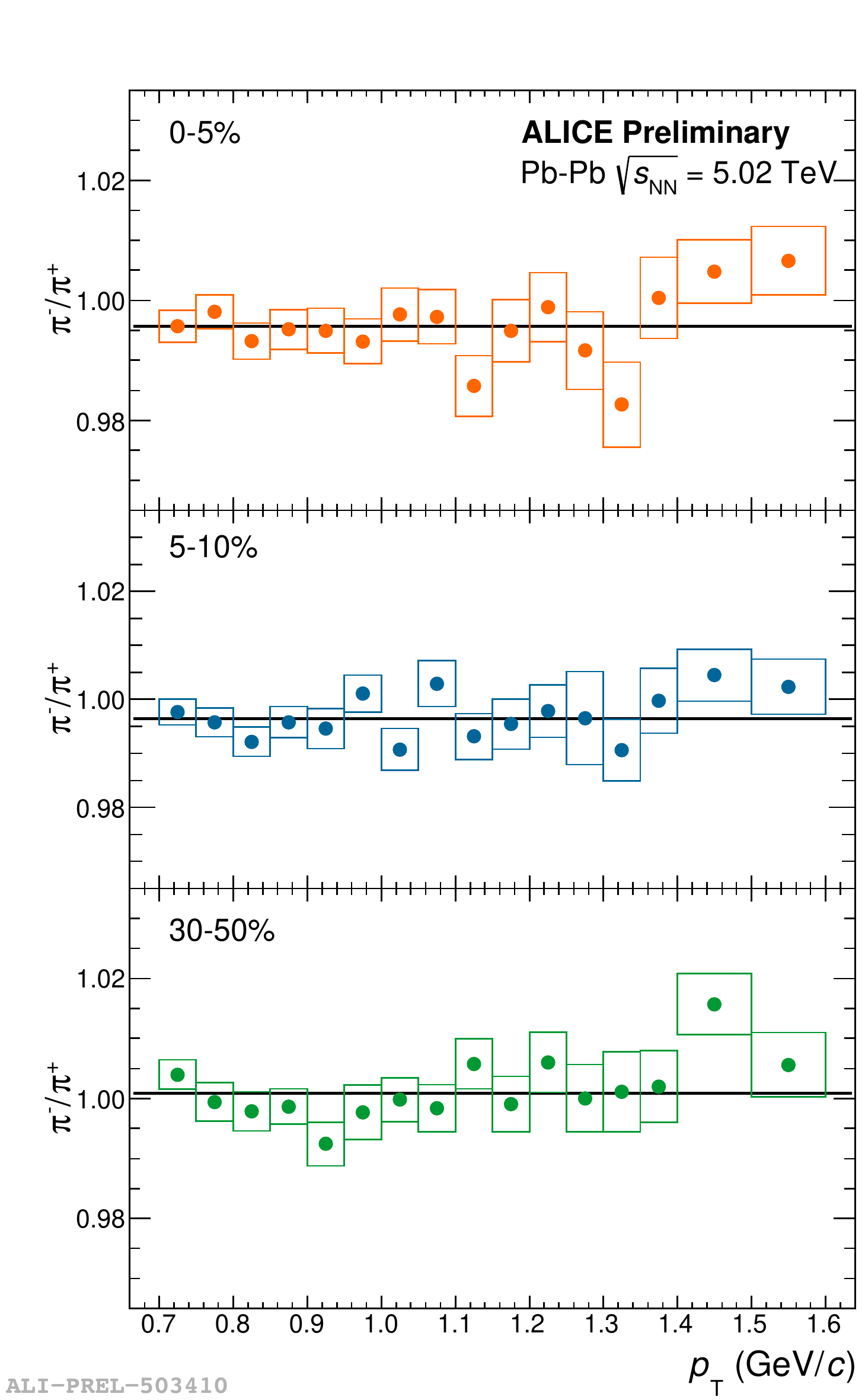}
\end{subfigure}
\begin{subfigure}{}
    \centering
    \includegraphics[height=0.6\textwidth,clip]{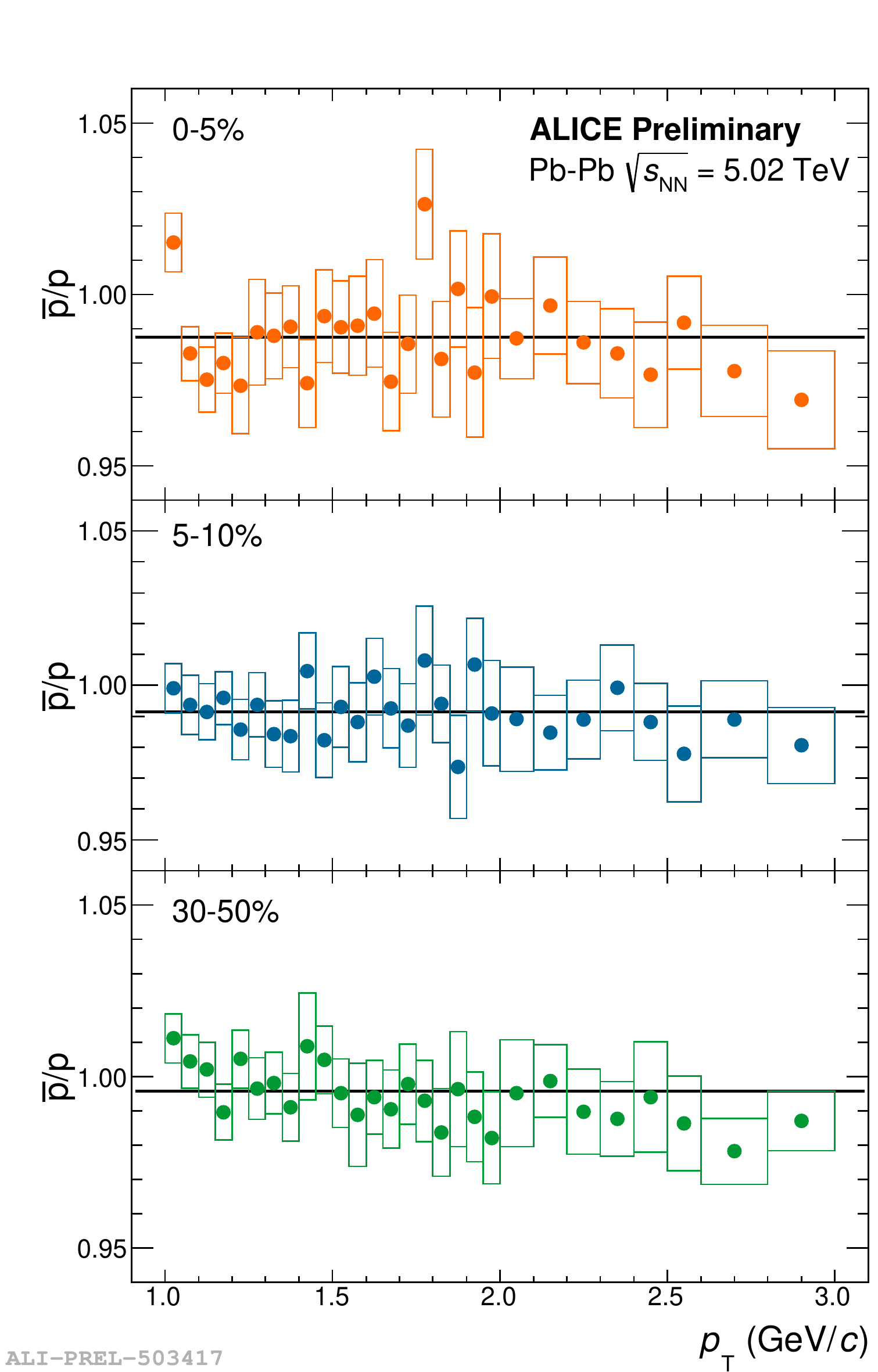}
\end{subfigure}
\begin{subfigure}{}
    \centering
    \includegraphics[height=0.6\textwidth,clip]{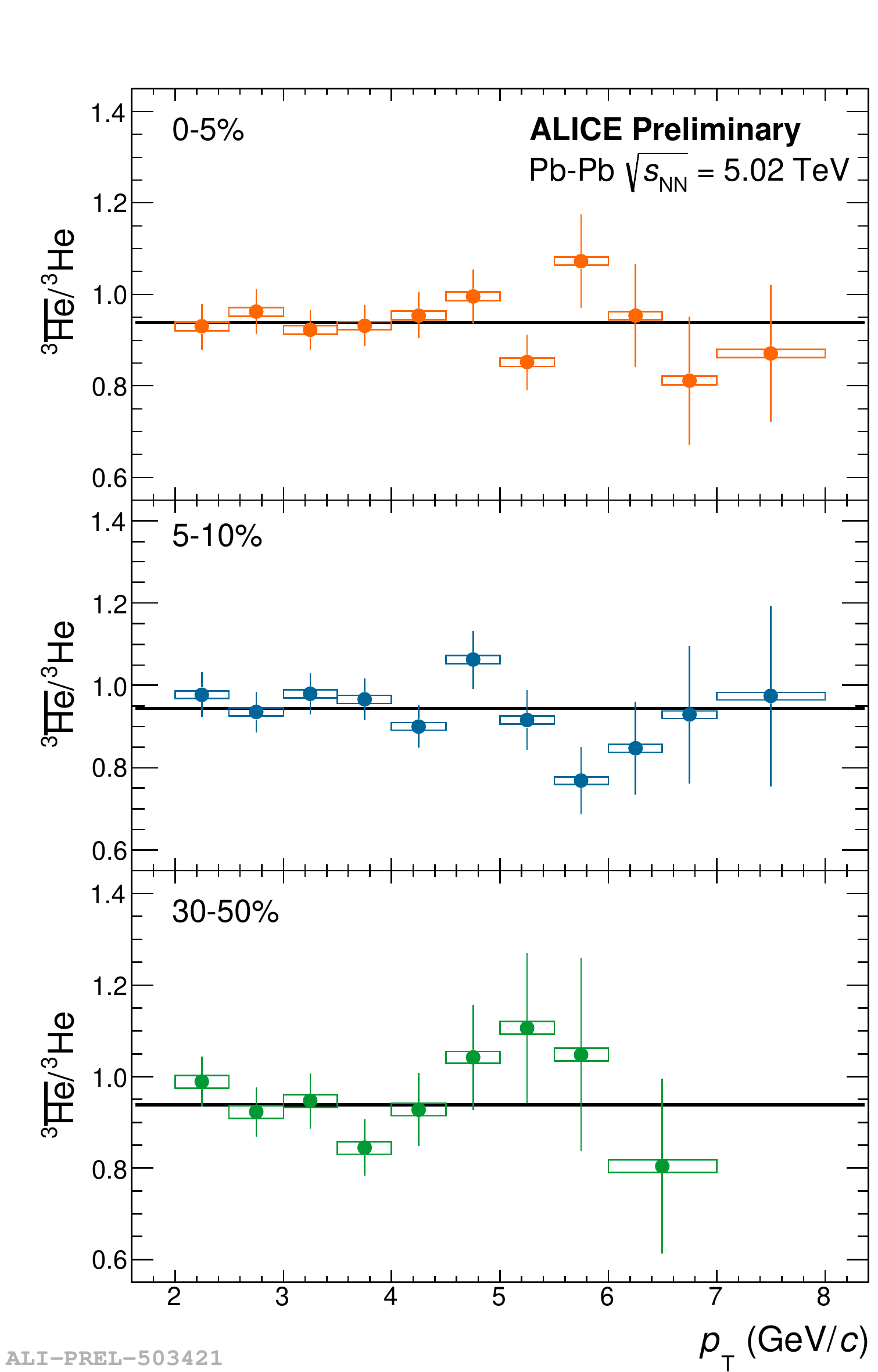}
\end{subfigure}
\begin{subfigure}
    \centering
    \includegraphics[height=0.6\textwidth,clip]{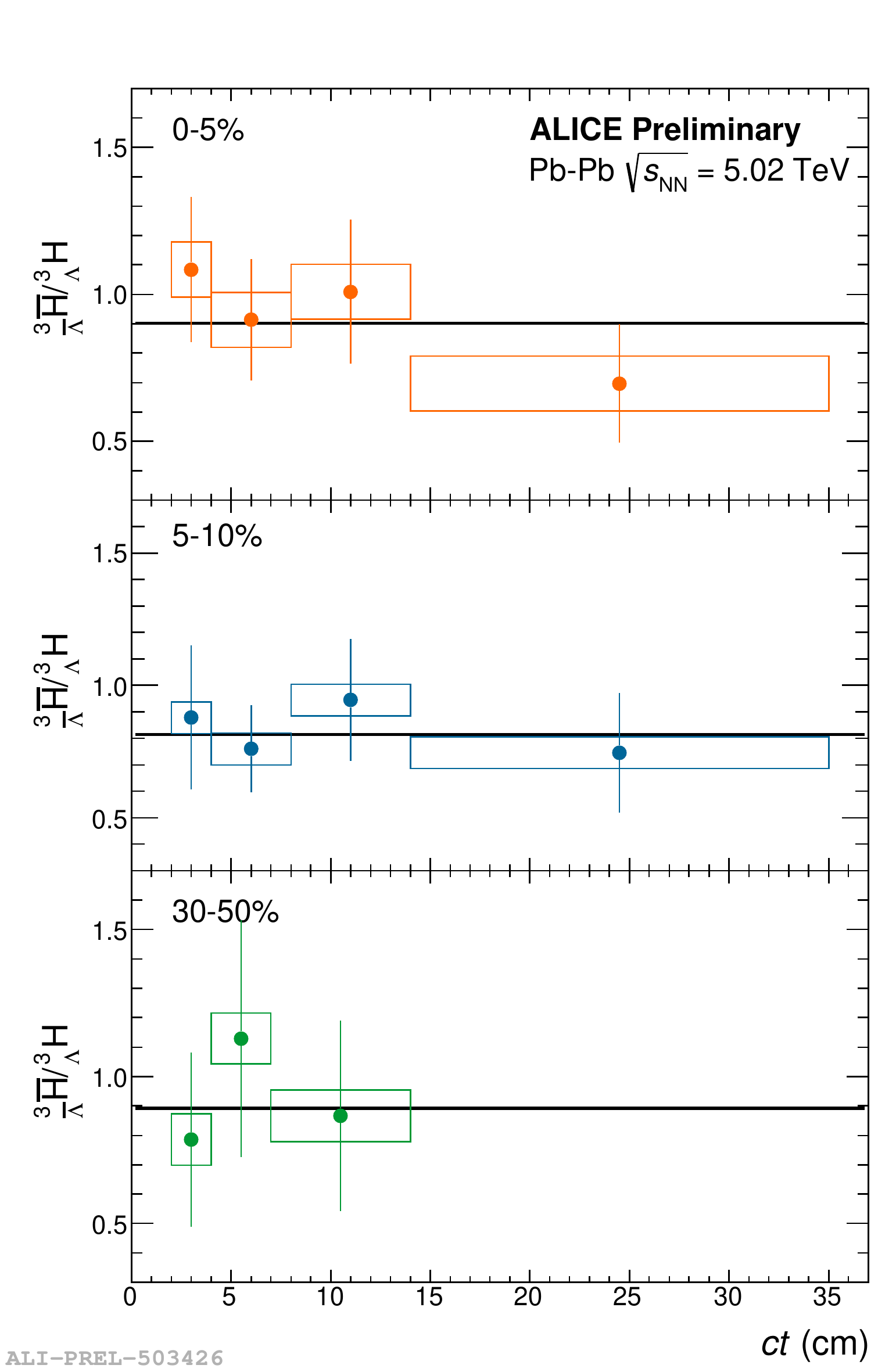}
\end{subfigure}
\caption{Antiparticle-to-particle ratios of charged pions (up, left), protons (up, right), ${}^{3}\mathrm{He}$ (low, left), in the 0-5\% (orange), 5-10\% (blue), and 30-50\% (green) centrality class, as a function of $p_{\mathrm{T}}$. The ratio as a function of $ct$ is shown for ${}^{3}_{\Lambda}\mathrm{H}$ (low, right). The error bars show the statistical uncertainty, while the boxes show the $p_{T}$ uncorrelated systematic uncertainty.}
\label{fig-1}       % Give a unique label
\end{figure}

The ratios of charged pions, protons, and helium are obtained as the ratios of the (anti)particle\footnote{The notation ``(anti)particle'' is used instead of ``antiparticle and particle'' for brevity.} yields in bins of transverse momentum $p_{\mathrm{T}}$, i.e. the projection of the track momentum on the plane transverse to the beam axis. The candidate track sample is obtained by applying selection conditions to the kinematic and reconstruction-quality variables of each track \cite{ref:ALICEpiKp, ref:ALICEnuclei}. In addition, the information on the distance of closest approach (DCA) of the backward extrapolation of the track to the primary vertex (PV) is used to reject candidates produced either in weak decays of multistrange baryons ($\pi^{\pm}$ and p) and hypertriton (${}^{3}\mathrm{He}$), or via spallation interactions of primary particles in the detector material. The yields are obtained by counting the number of reconstructed candidates that are compatible with the species of interest. For this purpose the PID information of the TPC is used to extract the ${}^{3}\mathrm{He}$ counts, as the $\mathrm{d}E/\mathrm{d}x$ of helium ($Z=2$) is four times larger than that of other particles of unitary electric charge at equal relativistic $\beta\gamma$. The considered transverse momentum range is $2\leq p_{\mathrm{T}}<8\ \mathrm{GeV}/c$. On the other hand, the $\pi^{\pm}$ and the (anti)proton yield is obtained from the TOF signal distribution after a preliminary selection based on the TPC $\mathrm{d}E/\mathrm{d}x$, in the range $0.7\leq p_{\mathrm{T}} < 1.6\ \mathrm{GeV}/c$ for pions and $1\leq p_{\mathrm{T}}<3\ \mathrm{GeV}/c$ for protons. The extracted yields are then corrected for the track reconstruction and selection efficiency, and by the geometrical acceptance of the detector, $\epsilon\times A$: this factor is computed from a detailed Monte Carlo simulation of the experiment. Moreover, the residual contamination due to secondary particles is estimated and subtracted from the raw yields. Systematic uncertainties on the extracted ratios are obtained by varying the track selection criteria and the signal extraction methods: the same variations are applied simultaneously both to the particle and corresponding antiparticle samples, thus cancelling out systematic effects that are correlated between charge-conjugated tracks. In addition, the uncertainty on the absorption cross section in the detector material and the uncertainty on the material budget itself are taken into account as centrality-correlated sources of systematic uncertainty. The results obtained for the charged pions, protons, and helium in each centrality interval are reported in Fig.~\ref{fig-1}. The weighted average of the $p_{\mathrm{T}}$-differential points is computed in each centrality interval: it is shown in black in the plots of Fig.~\ref{fig-1}.

The hypertriton is not directly tracked by ALICE due to the lack of enough space points, as it weakly decays with $c\tau\approx 8\ \mathrm{cm}$. Instead, it is reconstructed via its two-body charged-$\pi$ mesonic decay:
\begin{linenomath}
\begin{equation*}
{}^{3}_{\Lambda}\mathrm{H}\to {}^{3}\mathrm{He}+\pi^{-}\ (+\mathrm{c.c.})
\end{equation*}
\end{linenomath}
This decay channel is characterised by a branching ratio $\mathrm{B.R.=25\%}$ according to theoretical calculations \cite{ref:KamadaHyp} and by the clean identification of the produced ${}^{3}\mathrm{He}$ for the aforementioned reasons. The sample of candidates obtained by pairing tracks compatible with the helium and pion hypothesis is contaminated by a large fraction of combinatorial background, i.e. fake candidates produced by the association of uncorrelated $\pi$ and $^{3}\mathrm{He}$ candidates, since the pion yield is about $10^{6}\times$ larger than the ${}^{3}\mathrm{He}$ one in Pb-Pb collisions at the LHC. For this reason, machine-learning based selections are applied on top of loose standard preselections to enhance the ${}^{3}_{\Lambda}\mathrm{H}$ signal-to-background ratio     \cite{ref:ALICEHyppPb, ref:ALICEHypPbPb}. In particular, Boosted Decision Trees (BDT) are trained and tested on a sample of simulated hypertriton signal and data-driven background in bins of its proper decay length $ct$, using kinematic and topological variables both of the decay products and of the reconstructed decay vertex as training features. The trained BDTs are then applied to the data sample to extract the true ${}^{3}_{\Lambda}\mathrm{H}$ yield. The efficiency of the preliminary and BDT selections is computed using the simulated signal to correct the obtained counts. The results obtained in each centrality class are reported in the rightmost lower panel of Fig.~\ref{fig-1} as a function of $ct$. As for the other species, the weighted average of the measured points is reported in black.

\section{Results}
\begin{figure}[!]
\centering
\includegraphics[width=\textwidth,clip]{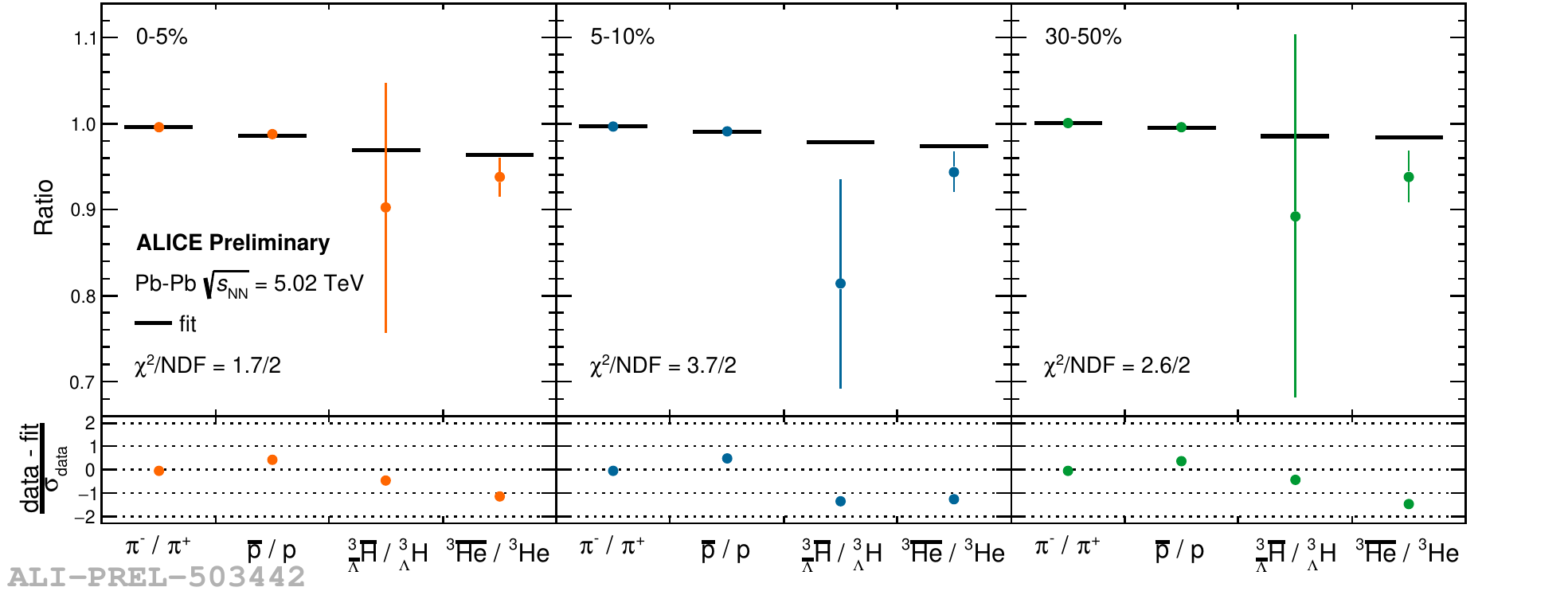}
\caption{Upper panel: Statistical Hadronisation Model fit to the antiparticle-to-particle ratios in the 0-5\% (orange), 5-10\% (blue), and 30-50\% (green) centrality classes. The fit results are reported in black. Lower panel: difference between data points and fit results, normalised to the uncertainty on the data.}
\label{fig-2}       % Give a unique label
\end{figure}

The baryon chemical potential is obtained by fitting the measured ratios with the relation based on the SHM relation reported in Eq.~\ref{eq-1}. In this equation, the baryon and isospin chemical potential $\mu_B$ and $\mu_{I_{3}}$ are the fit parameters, while the temperature is fixed to $T_{ch}=156.2\pm1.5\ \mathrm{MeV}$ \cite{ref:DecodingQCD}. The quantum numbers of the considered species appearing in Eq.~\ref{eq-1} are shown in Table~\ref{tab-1}: the linear combination $B+S/3$ is explicitly reported. The results obtained in the three analysed centrality classes are shown in Fig.~\ref{fig-2}. The uncertainty on the data points that are used for the fit is given by the sum in quadrature of the statistical and uncorrelated systematic uncertainty; the correlated part of the systematic uncertainty is estimated separately. A good fit quality is obtained in all of the analysed centrality intervals, as it can be observed both from the $\chi^{2}/\mathrm{NDF}$ and from the direct comparison of data and fit shown in the lower panel of Fig.~\ref{fig-2}. Moreover, it can be observed that the experimental data exhibit the same baryon-number hierarchy as predicted by the SHM.
\begin{table}[h]
    \caption{Quantum numbers of the analysed particle species.}
    \label{tab-1}
    \begin{tabular}{c c c c c} \hline           
         & $\pi^+$ & $\mathrm{p}$ & $^3\mathrm{He}$ & $^3_\Lambda\mathrm{H}$\\ \hline
        $B+S/3$ & 0 & 1 & 3 & 8/3 \\
        $I_3$ & 1 & 1/2 & 1/2 & 0 \\ \hline
    \end{tabular}
\end{table}

\begin{figure}[!]
\centering
\includegraphics[width=0.7\textwidth,clip]{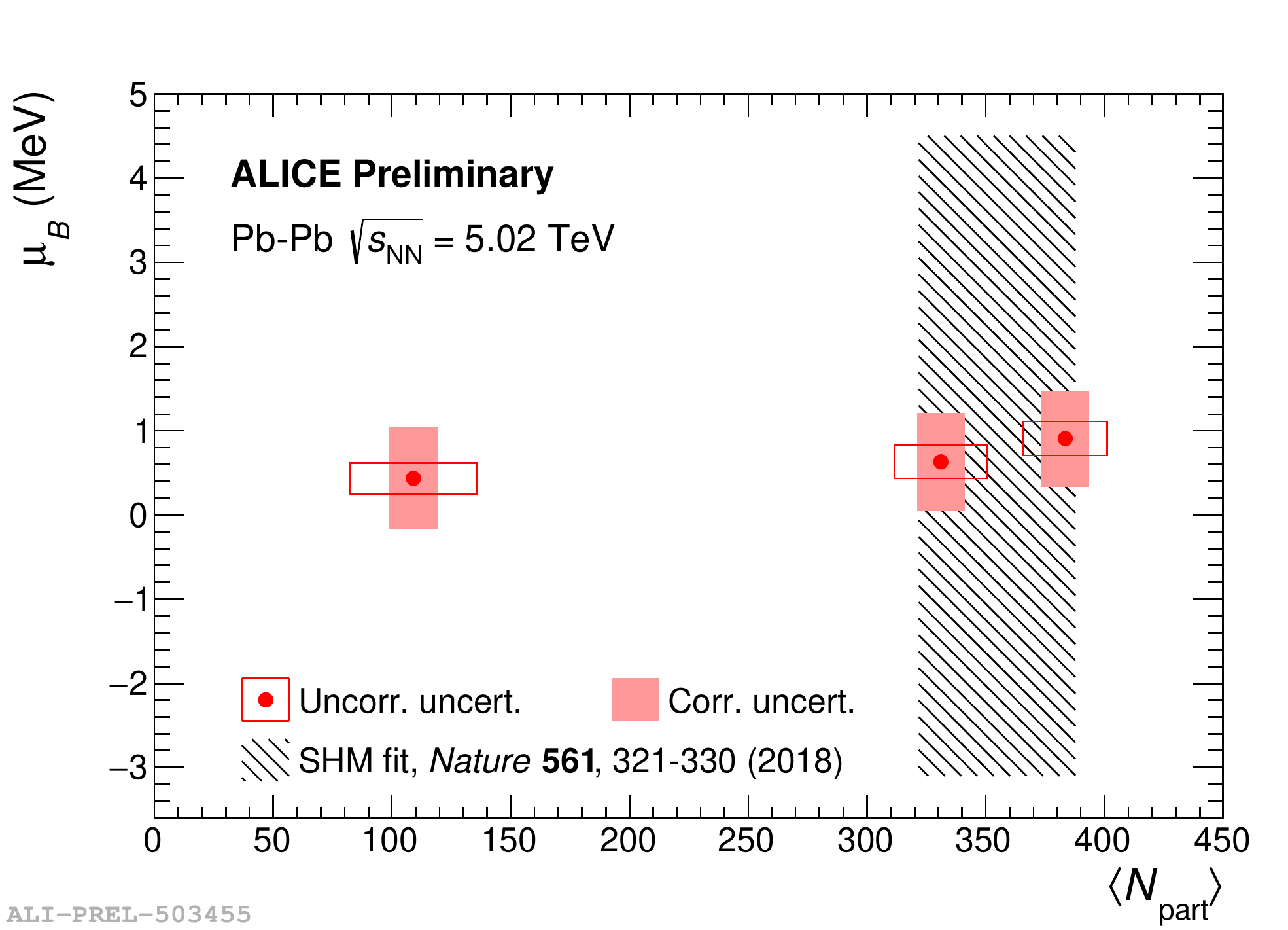}
\caption{Baryon chemical potential $\mu_B$ as a function of the average number of participating nucleons $\langle N_{\mathrm{part}}\rangle$. The open boxes show the uncorrelated uncertainties, while the shaded ones show the correlated ones. The previous measurement is shown as a black shaded area \cite{ref:DecodingQCD}.}
\label{fig-3}       % Give a unique label
\end{figure}

The $\mu_B$ obtained from the SHM fits to the ratios are shown in Fig.~\ref{fig-3} as a function of the average number of participating nucleons $\langle N_{\mathrm{part}}\rangle$, which is linked to the size of the interaction region, and consequently to the centrality of the collision. The first published $\mu_B$ measurement at the LHC \cite{ref:DecodingQCD} is also reported in this plot. It can be observed that the results obtained in this work are compatible with the previous one, while showing an improvement in the precision by about one order of magnitude due to the larger data sample and different analysis technique. Since three different centrality intervals are considered, it is also possible to study a possible centrality dependence of $\mu_B$, which could be caused by a decrease in the baryon stopping when going from the most central to semicentral events. No evidence of such a dependence is found within the current precision.

\section{Conclusions}
In this contribution, the most precise centrality-differential $\mu_B$ measurement available in Pb-Pb collisions at the LHC is reported. This result is achieved by studying the antiparticle-to-particle ratios of various hadron species within the SHM picture, rather than their (anti)particle yields. Besides a significant improvement in the precision, this measurement allows us to exclude a centrality dependence of $\mu_B$. This study will be extended with the test of the isospin and strangeness dependence of ratios by considering also the triton $^{3}\mathrm{H}$, which is the isospin-symmetric counterpart of $^{3}\mathrm{He}$, and multistrange baryons such as the $\Omega$, for which a compensation between the baryon and strangeness hierarchy of the ratios is expected.

\end{document}